\documentclass[fleqn,usenatbib]{mnras}
\usepackage{hyperref}
\hypersetup{
    colorlinks=true,
    linkcolor=blue,
    filecolor=magenta,      
    urlcolor=cyan,
    pdfpagemode=FullScreen,
    }

\usepackage{newtxtext,newtxmath}
\usepackage[T1]{fontenc}
\usepackage{ae,aecompl}
\usepackage{graphicx}
\usepackage{amsmath}
\usepackage{lipsum}

\usepackage{multirow}
\usepackage{dcolumn}
\usepackage{soul}

\title[Heavy seeds as outliers in black hole vs. host galaxy relations]{How long do high-redshift massive black hole seeds remain outliers in black hole vs. host galaxy relations?}

\author[M.~T.~Scoggins et al]{Matthew~T.~Scoggins,$^{1}$\thanks{E-mail: mts2188@columbia.edu}
Zolt{\'{a}}n~Haiman,$^{1}$\ and
John~H.~Wise$^{2}$ 
\\
$^{1}$Department of Astronomy, Columbia University, New York, NY, 10027\\
$^{2}$Center for Relativistic Astrophysics, School of Physics, Georgia Institute of Technology, 837 State Street, Atlanta, GA 30332\\
}

\pubyear{2022}

\begin{document}
\label{firstpage}
\pagerange{\pageref{firstpage}--\pageref{lastpage}}
\maketitle

\begin{abstract}
    The existence of $10^9\ {\rm M_\odot}$ supermassive black holes (SMBHs) within the first billion years of the universe remains a puzzle in our conventional understanding of black hole formation and growth. Several suggested formation pathways for these SMBHs lead to a heavy seed, with an initial black hole mass of $10^4-10^6~{\rm M_\odot}$. This can lead to an overly massive BH galaxy (OMBG), whose nuclear black hole's mass is comparable to or even greater than the surrounding stellar mass: the black hole to stellar mass ratio is $M_{\rm bh}/M_* \gg 10^{-3}$, well in excess of the typical values at lower redshift. We investigate how long these newborn BHs remain outliers in the $M_{\rm bh}-M_{*}$ relation, by exploring the subsequent evolution of two OMBGs previously identified in the \texttt{Renaissance} simulations. We find that both OMBGs have $M_{\rm bh}/M_* > 1$ during their entire life, from their birth at $z\approx 15$ until they merge with much more massive haloes at $z\approx 8$. We find that the OMBGs are spatially resolvable from their more massive, $10^{11}~{\rm M_\odot}$, neighboring haloes until their mergers are complete at $z\approx 8$. This affords a window for future observations with {\it JWST} and sensitive X-ray telescopes to diagnose the heavy-seed scenario, by detecting similar OMBGs and establishing their uniquely high black hole-to-stellar mass ratio.
\end{abstract}

\section{Introduction}

The origin of supermassive black holes (SMBHs) larger than $10^9~{\rm M_\odot}$ powering quasars at redshifts $z\geq 6$ remains poorly understood. There are over 200 detections of these SMBHs (for recent compilations, see \citealt{Inayoshi_2020} and \citealt{Bosman_2022})
with likely many more below the current observational threshold. The existence of black holes of $\sim10^9 \ {\rm M_\odot}$ before the first billion years of the universe requires adjustments to our current understanding of black hole formation and growth.

Several scenarios have emerged which attempt to explain their existence (for recent reviews, see \citealt{Inayoshi_2020} and \citealt{Volonteri_2021}). The "light seed" scenario (e.g. \citealt{Tanaka_2009,Volonteri_2010}) proposes a Population III (Pop III) metal-free star (e.g. \citealt{Abel_2000, Bromm_2001, Abel_2002, Yoshida_2008, Clark_2011, Greif_2011, Hirano_2014, Susa_2014, Stacy_2016}) of $\sim10$-$100 \ {\rm M_\odot}$ which forms the SMBH seed. A $\sim 100 \ {\rm M_\odot}$ Pop III seed would need to grow near the Eddington limit, uninterrupted for the age of the $z \sim 6$ universe, to explain the SMBHs at high redshift. However, these Pop III remnants are expected to be born in warm, diffuse regions which prevent their growth \citep{Whalen_2004}. Once incorporated into a galaxy, they are still typically located in underdense, off-centre regions, leading to accretion at orders of magnitude below the Eddington rate (\citealt{Alvarez_2009, Milosavljevi__2009, Tanaka_2009, Tanaka_2012,Smith_2018, Regan_2019,Pfister_2019, Regan_2020b}). Growth via mergers is also hindered by gravitational-wave recoil, which often ejects black holes from the shallow potentials of their host haloes \citep{Haiman_2004}. Periods of mildly super-Eddington accretion (e.g. \citealt{Madau_2001, Madau_2014, Alexander_2014, Lupi_2016}) and/or short periods of hyper-Eddington accretion (e.g. \citealt{Inayoshi_2016, Pacucci_2017,Hu_2022a}) could explain the rapid growth of light seeds, but it remains unclear how often this accelerated growth is realized in nature.

An alternative pathway, the "heavy seed" scenario, proposes black hole seeds which start with $\sim10^4-10^6~{\rm M_\odot}$. One of the most studied versions of the heavy seed pathway is the so-called direct-collapse black hole (DCBH), where a high accretion rate onto protostars allows for the creation of a short-lived supermassive star (SMS) that leads to the $\sim10^4-10^6~{\rm M_\odot}$ seed. The formation of these DCBHs (e.g. \citealt{Omukai_2001, Oh_2002, Bromm_2003, Begelman_2006, Spaans_2006, Shang_2010, Agarwal_2012, Hosokawa_2012, Hosokawa_2016, Latif_2013, Ferrara_2014, Inayoshi_2014, Sugimura_2014, Tanaka_2014, Becerra_2015, Chon_2016, Umeda_2016, Hirano_2017, Haemmerle_2018}), are believed to require special environments in chemically pristine atomic-cooling haloes which allow rapid collapse and formation of a SMS. It has recently been shown that the metal-free condition is not strictly necessary. Extremely metal poor haloes with $Z \ < \ 10^{-3}~{\rm Z}_\odot$ can still allow the rapid growth of a central protostar at $\sim 1~{\rm M_\odot}~{\rm yr}^{-1}$, leading to a SMS of $\sim 10^5\ {\rm M_\odot}$ \citep{Tagawa_2020, Chon_2020, Regan_2020}.

Rapid central collapse can be achieved through intense Lyman-Werner (LW) radiation from a neighboring galaxy (suppressing ${\rm H_2}$-cooling), dynamical heating from rapid halo mergers (increasing the heating rate), or large residual baryonic streaming motions from recombination (preventing gas infall and contraction into DM haloes). These processes, or some combination of them, are invoked to suppress ${\rm H_2}$ formation, cooling of the gas, and star formation, until the haloes cross the so-called atomic cooling threshold, with masses of $10^7-10^8~{\rm M_\odot}$.  Once this threshold is crossed, atomic hydrogen cooling can result in the catastrophic collapse of the halo's pristine gas.  As a result of hosting little to no prior star formation, massive seeds are generally believed to form in relatively small haloes, containing no or very few stars (a feature emphasized by, e.g. \citealt{Agarwal_2013}). Here we dub these objects "Overly Massive Black Hole Galaxies" (OMBGs). The mass of the black hole dominates the halo's total stellar mass $M_*$, with $M_{\rm bh}/M_* \gg 10^{-3}$, where $\sim10^{-3}$ is the typical mass ratio for galaxies at lower redshift~\citep{Sani_2011,KormendyHo2013}.

There are several other scenarios which could lead to the rapid formation of a heavy seed. This includes massive primordial star clusters where collisions can give rise to seeds of up $10^4-10^5 {\rm M}_\odot$ \citep{Boekholt_2018,Tagawa_2020, Escala_2021, Vergara_2022, Schleicher_2022}. It is also physically viable that a small black hole can quickly become a $ 10^5\ {\rm M_\odot}$ seed via hyper-Eddington accretion \citep{Ryu_2016,Inayoshi_2016}. These scenarios usually involve fragmentation and star formation prior to forming the massive seed, so it is not clear how $M_{\rm bh}/M_*$ relation holds over time. However, given that all of the above scenarios occur in ACHs and produce an initially heavy seed BH with comparable mass and a dearth of accompanying stars, we expect the subsequent $M_{\rm bh}/M_*$ relations to be similar. Therefore, using this mass ratio as a probe would also apply to these other formation pathways.

In this paper, we consider the heavy seed  scenario and explore how long massive DCBH seeds remain outliers in the $M_{\rm bh}$ {\it vs.} $M_*$ relation. This is a key question for attempts to diagnose the heavy seed formation pathways via this distinguishing feature. The question was considered recently in \citet{Visbal_2018}, but only for black holes born in random atomic-cooling haloes, not accounting for the bias that occurs when requiring that the parent atomic-cooling halo should end up as the massive host of a high-redshift quasar at $z=6-7$~\citep{Lupi_2021, Li_2021}. Here we consider two specific OMBGs identified in the \texttt{Renaissance} simulations by \citet[][hereafter W19]{Wise_2019}, and follow the subsequent evolution of $M_{\rm bh}$ and $M_*$ in these two haloes. If the $M_{\rm bh}/M_*$ ratio remains abnormally large through redshift $z=10$ or later, and SMBH hosts are resolvable, then this formation mechanism might be possible to corroborate through direct detection with a combination of infrared (e.g. {\it JWST}) and X-ray (e.g. {\it Athena} or {\it Lynx}) observations. W19 identified the direct-collapse scenario as a possible outcome of rapid mass inflow in pristine ACHs in their simulations. However, we emphasize that our results, stated for DCBHs throughout this paper, would also apply to massive seed BHs forming in these haloes through either of the other heavy-seed pathways.

The rest of this paper is organised as follows. In \S~\ref{sec:methods} we describe the \texttt{Renaissance} simulation data, our target OMBGs, the process of halo finding and creating merger histories, and our modeling of the black hole and stellar masses during periods of growth and tidal stripping for our OMBGs. In \S~\ref{sec:results_and_discussion} we present and discuss our results for black hole mass, stellar mass, and $M_{\rm bh}/M_*$. Specifically, we note that across a variety of parameters that govern the growth of our black holes, this unique mass relation stays well above the lower-redshift value up to $z=8$, and possibly further. We go on to compare our value of  $M_{\rm bh}/M_*$ to the value expected in light-seed pathways, consider an alternative stellar-mass calculation, and alternative model for BH growth, and discuss the DCBH number density and the possibility of detection. We also consider other probes which would distinguish heavy-seed {\it vs.} light-seed pathways. Finally, we summarize our findings and offer our~conclusions in~\S~\ref{sec:conclusion}. Our analysis and data used in this work assume the following cosmological parameters: $\Omega_{\Lambda} = 0.734$, $\Omega_m = 0.266$, $\Omega_b=0.049$, and $h =0.71$.

\section{Methods}
\label{sec:methods}

\subsection{Our target DCBH-hosting haloes, MMH \& LWH}
Our work focuses on two target DCBH-hosting haloes previously identified by W19, where they perform a suite of cosmological radiation-hydrodynamic and N-body simulations, dubbed \texttt{Renaissance} \citep{OShea_2015, Xu_2016}, with the adaptive mesh refinement code \texttt{Enzo} \citep{Enzo, Enzo_2019}. \texttt{Renaissance} is divided intro three regions of high ($\langle \delta \rangle \equiv \langle \rho \rangle /(\Omega_{\rm M} \rho_{\rm c} - 1 \sim 0.68$), average ($\langle \delta \rangle \sim 0.09$), and low  ($\langle \delta \rangle \sim -0.26$) mean overdensity, referred to, respectively, as the {\it Rarepeak},  {\it Normal}, and {\it Void} regions.  Inspecting the 133.6 (comoving) ${\rm Mpc}^3$ {\it Rarepeak} region yields 11 metal-free atomic cooling haloes which have not hosted star formation prior to $z = 15$. Of these 11 candidate DCBH haloes, two target haloes are identified: the most massive halo (MMH) and the most irradiated halo (LWH) which sees the highest Lyman-Werner flux. These targets are then re-simulated with a mass resolution higher by a factor of 169. This re-simulation follows the evolution of these targets until a density of $10^{-15}$ g ${\rm cm}^{-3}$ is reached,  where a collapsed object will likely form. Both haloes form in a region $\sim 20$ kpc away from a group of young galaxies that have photo-ionized, photo-heated, and chemically enriched their adjacent environments. The chemically enriched regions only extend $\sim 5$ kpc away from their centres, far from reaching the target haloes. These massive star-forming regions intensely radiate the target haloes, with both experiencing $J_{\text{LW}} \sim 3 J_{21}$ at $z=15$, where $J_{\rm LW}$ is the intensity of the radiation at $\sim$ 12~eV, in units of $J_{21}=10^{-21}$ erg cm$^{-2}$ s$^{-1}$ Hz$^{-1}$ sr$^{-1}$.The total flux on the target haloes is 6--600 times lower than previous estimates of the critical value for SMS formation \citep{Shang_2010, Agarwal_2016, Glover_2015, Wolcott-Green+2017}, meaning the high mass infall rates must be achieved through other means. W19 find that dynamical heating via mergers plays the primary role in preventing the formation of Pop III stars and allowing the formation of a SMS.

The original hydrodynamic simulations in the {\it Rarepeak} region of \texttt{Renaissance} only run to redshift $z=15$. With the goal of measuring $M_{\rm bh}/M_*$ down to a redshift visible by the JWST ($z\lesssim 10$), we use a corresponding dark matter (DM) only N-body simulation which extends down to redshift $z=6$.  This DM simulation uses the same initial conditions and mass resolution as the hydrodynamical simulations in the \texttt{Renaissance} suite, but has lower spatial resolution and extends to lower redshift.

Using the \texttt{ROCKSTAR} halo finder \citep{rockstar} to identify haloes and \texttt{CONSISTENT$\_$TREES} \citep{con_trees} to construct merger histories, we identify MMH and LWH in this DM simulation by using the coordinates and velocities of MMH and LWH in the \texttt{Renaissance} hydrodynamic simulation at $z=15$ and approximating the target halo positions at the closest snapshot in our DM simulation, at redshift $z=14.926$. A simple linear approximation of $x_{{\rm i,}z=14.92} \approx x_{{\rm i,}z=15} + v_{{\rm i,}z=15}dt$ is sufficient considering that the total time difference is $dt = 1.937$ Myr and each halo is moving at roughly $150$ km s$^{-1}$, meaning each halo travels no more than $0.5$ kpc, less than their virial radii of $\sim 1$ kpc\footnote{Unless stated otherwise, all distances in this paper are quoted in physical (not comoving) units.}. We search a $\sim 6$ ${\rm kpc}^3$ box centred on these approximated positions, and in each case find only one halo at each location with properties that match MMH and LWH, meaning we have successfully found our target haloes. The haloes identified at the $z=14.926$ positions have masses slightly smaller than the masses of MMH and LWH given in W19. The lower-resolution DM simulation likely fails to resolve several small mergers which were captured in the zoom-in hydrodynamical simulation in W19 where MMH and LWH were first identified.

\subsection{Calculating stellar and black hole mass}
\label{sec:methodsSM}

We assign stellar masses to our haloes following \citet{Behroozi_2019}, who use a combination of simulation data and observational constraints to fit median stellar mass to halo mass and redshift. Specifically, we adopt their Appendix~J with constants adopted from their Table~J1. Constants are chosen as a function of the haloes being: stellar mass (SM) being true or observed; star forming vs quenched (SF/Q); satellite or central haloes (Sat/Cen); and including or excluding intrahalo light (IHL). We choose row 15 of the table, corresponding to the true stellar mass for star forming central and satellite haloes. This only leaves the option to exclude IHL. (SM=True, SF/Q=SF, Sat/Cen=All, IHL=Excl). Equation~J1 in \citet{Behroozi_2019} comes from best-fitting the median ratio of stellar mass to peak historical halo mass ($M_{ \rm peak}$), the maximum mass attained over the halo's assembly history. These formulae were fit only for haloes with masses $10^{10.5} {\rm M}_\odot < M_{ \rm peak} < 10^{15} {\rm M}_\odot$, forcing us to  extrapolate to obtain approximate stellar masses below this range.

Stellar mass generally grows monotonically over time, with gradual increases corresponding to net star formation in the host haloes, and abrupt jumps corresponding to stellar mass acquired during mergers. There can be periods of stellar mass decline, either due to tidal stripping or natural mass loss through stellar winds and supernovae whenever the star-formation rate does not offset this loss. We note that the stellar masses we adopt account for all of these effects. Our haloes experience stellar mass decline during tidal stripping events, when our target haloes pass near or through more massive haloes. We expect the stripped fraction of stellar mass to be much smaller than the fraction of dark matter mass, due to the concentration of stellar mass near the centre of the halo \citep[e.g.][]{Smith_2016,Costa_2019}, and the stellar masses account for this.  In \S~\ref{sec:discussion_2}, we discuss an alternative stellar mass prescription, which uses instantaneous halo mass and carefully accounts for periods of tidal stripping.

Black holes are born at times determined in W19, roughly corresponding to when their host haloes cross the atomic-cooling threshold, with MMH's black hole born at $z=16.4$ and LWH's black hole born at $z=15.3$. The black holes' initial masses and growth are explored in detail in \citet{Inayoshi_2020}. For our purposes, we explore a range of parameters. Initial seed masses in the \texttt{Renaissance} simulation are estimated to fall within the range $10^4  {\rm M_\odot}\leq M_{\rm bh} \leq 10^6  {\rm M_\odot}$, in agreement with the expected seed mass for the DCBH formation pathway, so we explore an initial seed mass of $M_{\rm i} \in \{10^4, 10^5, 10^6\}$ ${\rm M_\odot}$. The growth rate is assumed to follow the Eddington rate 
\begin{align}
   & L_{\rm edd} \equiv \frac{4\pi c G \mu m_{\rm p} M_{\rm bh}}{\sigma_{\rm T}} = \epsilon c^2 \dot{M}_{\rm bh},
\end{align} 
with speed of light $c$, gravitational constant $G$, mean molecular weight $\mu$ ($\mu \sim 0.6$ for primordial ionized gas), proton mass $m_{\rm p}$, Thomson cross section $\sigma_{\rm T}$,  and radiative efficiency $\epsilon$. This leads to a black hole mass given by $M_{\rm bh}(t) = M_{\rm i}\exp(t/\tau_{\rm fold})$ with e-folding time $\tau_{\rm fold} =  (\sigma_{\rm T} c \epsilon)/(4\pi \mu G m_{\rm p}) \approx  $ 450$\epsilon$ Myr. Assuming efficiency $\epsilon \approx 0.1$ and allowing variations in $\epsilon$ due to BH spin, we consider $\tau_{\rm fold} \in \{20, 40, 80\}$ Myr. We additionally quench black hole growth when the mass of the black hole exceeds a prescribed fraction of the baryonic matter in the halo, capping $M_{\rm bh} \leq f_{\rm bh} M_{\rm halo} \Omega_{\rm b}/\Omega_{\rm m}$ with $f_{\rm bh}  \in \{ 0.05, 0.1, 0.2, 0.5\}$. To summarize, our simple model has three parameters: $M_{\rm i}, \tau_{\rm fold}$ and $f_{\rm bh}$.

\section{Results and Discussion}
\label{sec:results_and_discussion}

\begin{figure}
 \includegraphics[width=\columnwidth]{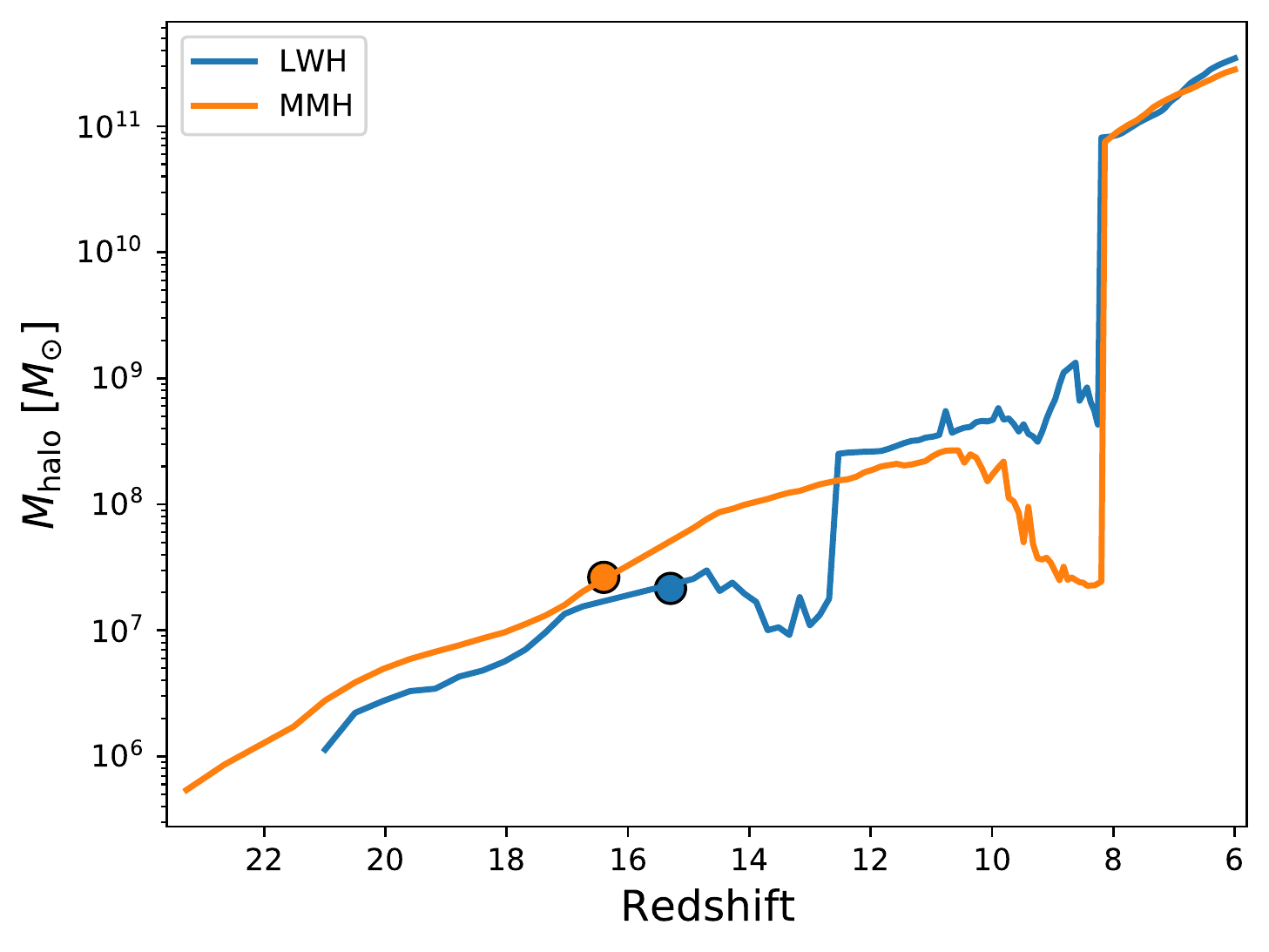}
 \caption{The total mass of our two target haloes, MMH and LWH, as a function of redshift. The formation time of the black hole in each halo is marked by the dots; $z=16.4$ and $z=15.3$ in MMH and LWH, respectively. MMH experiences tidal stripping from $z=11-8$ as it passes through a more massive "Superhost" halo, eventually completely merging with it at redshift $z\sim8$. LWH also experiences a smaller tidal stripping event near $z\sim14$, then merging completely with its Superhost at redshift $z\sim8$. The massive haloes that MMH and LWH merge with are two distinct haloes, they are $>2$ Mpc apart at redshift $z>6$.
 }
 \label{fig:halo_mass_vs_z_full}
\end{figure}

\begin{figure*}
 \includegraphics[width=0.9\textwidth, height=0.65\textheight]{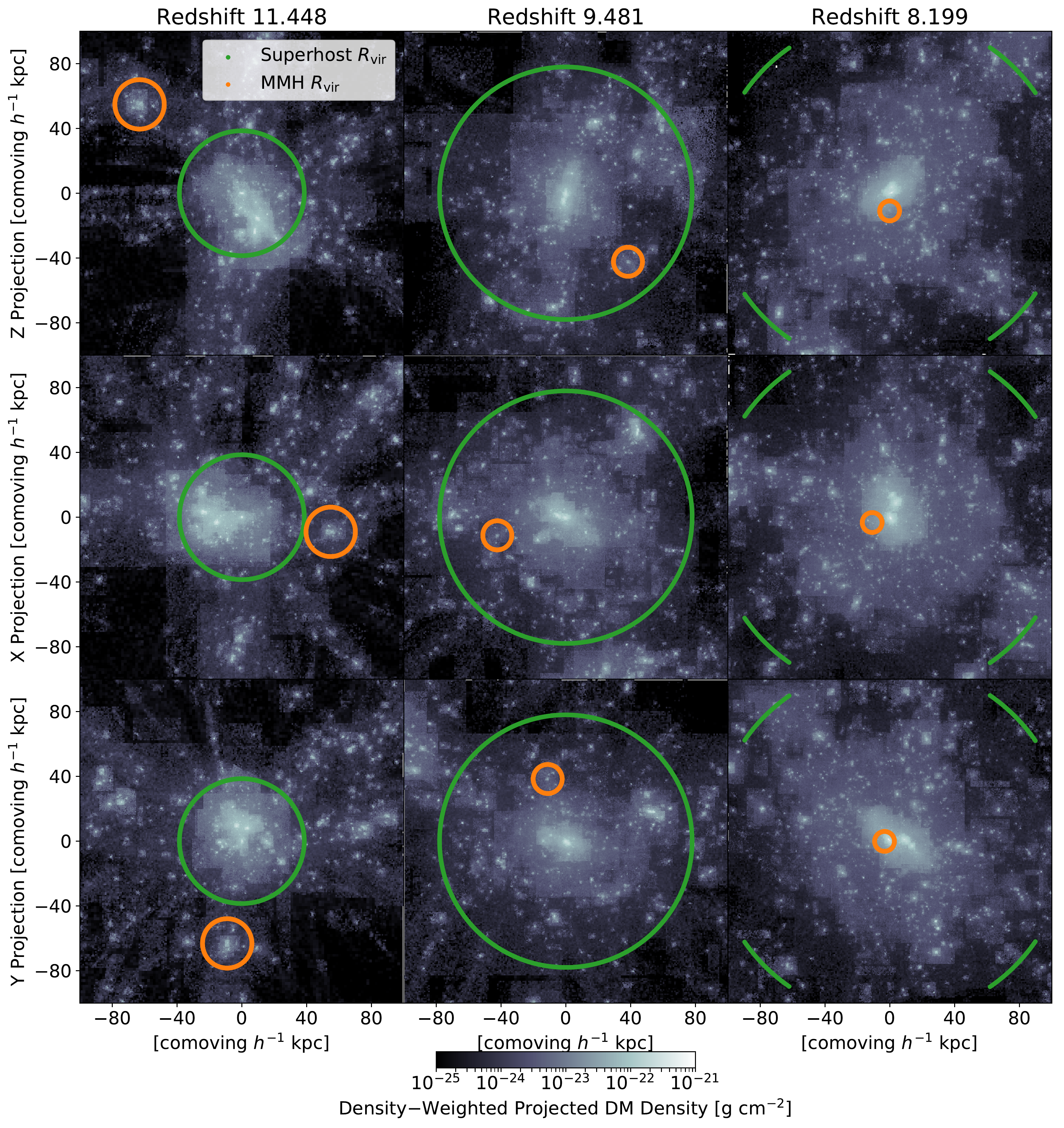}
 \includegraphics[width=0.9\textwidth, height=0.25\textheight]{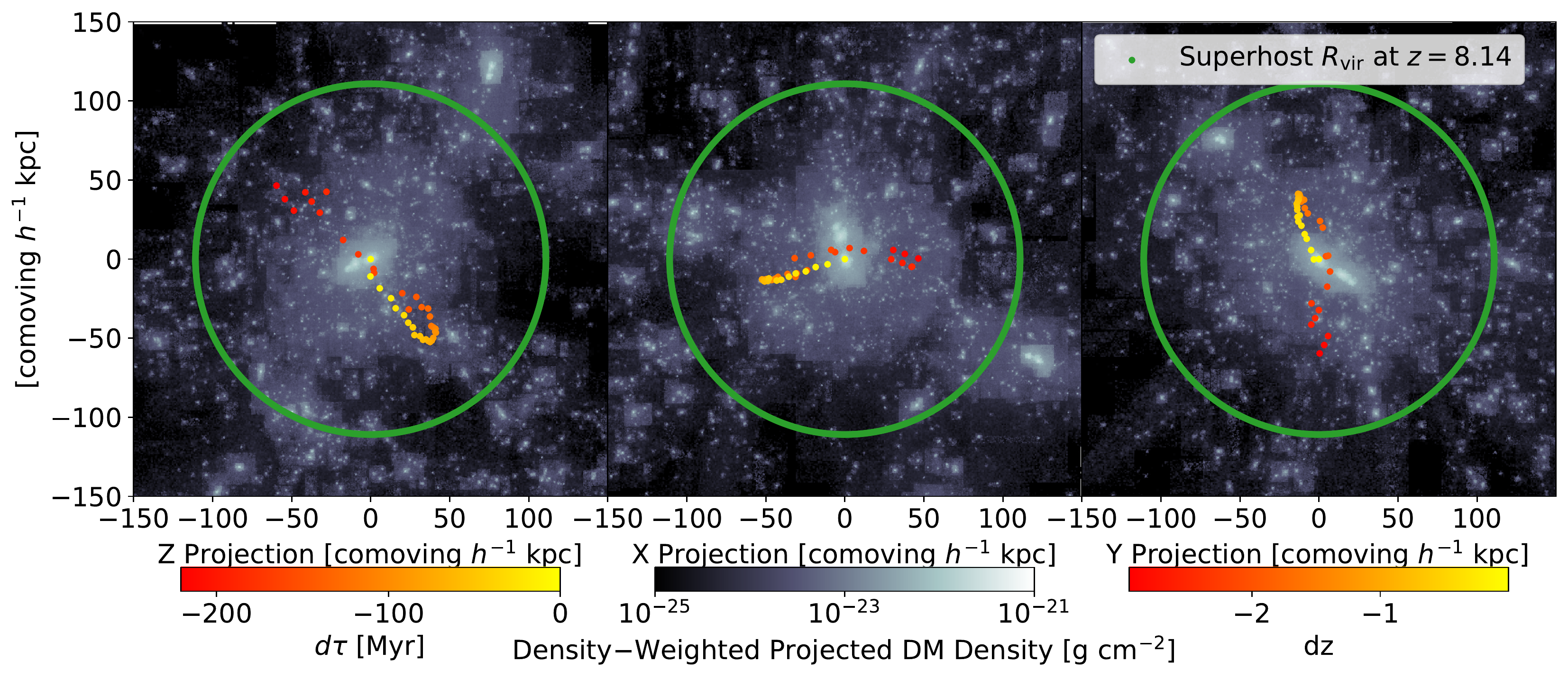}
 \caption{{\it Top:} 
The projected densities along the X, Y, and Z axes of MMH and its massive "Superhost" neighbor. Plots are centred on the Superhost, with columns showing redshift $z \in \{11.45, 9.48, 8.12\}$. Distances are in units of (comoving) $h^{-1}$kpc. Green circles show the virial radius of the Superhost. MMH becomes a subhalo of its Superhost near redshift $z\sim10.7$, when it begins losing mass due to tidal stripping. MMH passes near the centre of the halo near redshift $z\sim10$ then reaches its furthest distance at redshift $z\sim9$ before completing the merger at redshift $z=8.14$. The full movie of this collision is available at \href{https://mscoggs.github.io/visualizations_tools.html}{MMH-Collision}. 
{\it Bottom:} The centre-of-mass separation of MMH and its Superhost in the Superhost's frame at redshift $z=8.14$, when the merger is complete and MMH is no longer distinguishable from the Superhost. The dots show the separation between the two haloes as a function of time, with earlier times marked in red and later times marked in yellow.}
 \label{fig:collision}
\end{figure*}

Fig.~\ref{fig:halo_mass_vs_z_full} shows the halo mass histories of both targets, with the formation time of the black hole in each halo marked by dots; $z=16.4$ and $z=15.3$ in MMH and LWH, respectively. MMH experiences a period of mass loss due to tidal stripping from redshift $z\sim11$ to redshift $z\sim8$. The MMH merges with and passes near the centre of a more massive halo until it is no longer distinguishable from this massive halo at redshift $z=8.14$. We refer to a more massive halo which a DCBH-hosting target halo eventually merges with as a "Superhost". (The DCBH-host halo becomes, temporarily, a subhalo of this Superhost.) Snapshots of this flyby are shown in the top panel of Fig.~\ref{fig:collision}, which includes X, Y, and Z projection plots for our MMH at redshift $z \in \{11.45, 9.48, 8.2\}$, centred on MMH's Superhost, with the Superhost's virial radius in green and the virial radius of the MMH in orange.  Fig.~\ref{fig:collision} also shows the relative centre of mass (COM) separation (dots) during this close-encounter, in the frame of the Superhost at redshift $z=8.14$, with earlier times marked by red and later times marked yellow. LWH also temporarily merges with a Superhost before merging completely at redshift $z=8.256$. The similar timing of these mergers is a coincidence -- the two target haloes' Superhosts are distinct and independent; they are separated by $>2$ Mpc at redshifts $z>6$.

\begin{figure}
 \includegraphics[width=\columnwidth]{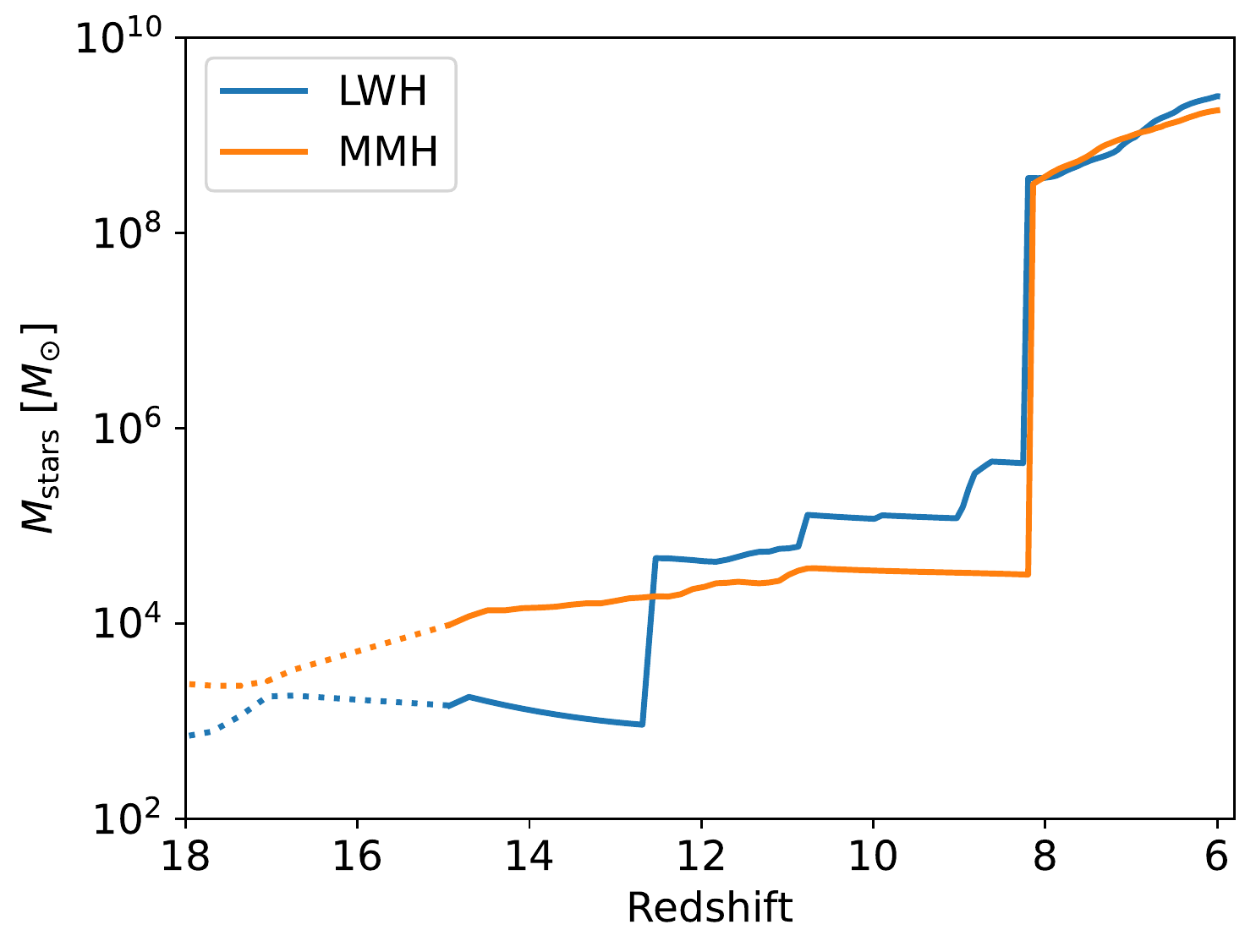}
 \caption{Stellar mass {\it vs.} redshift, converting $M_{\text{peak}}$, the peak historical halo mass, to stellar mass using the function introduced in Appendix~J of \citet{Behroozi_2019}. The stellar mass typically gradually increases (star formation) or sharply increases (mergers with star-hosting haloes), but occasionally declines due to natural stellar mass loss via stellar winds or tidal stripping. MMH loses stellar mass due to tidal stripping during a flyby with its Superhost from $z\sim11-8$. This stellar mass is lost at a rate which is much smaller than halo mass loss rate, as the stars would concentrate near the centre of the dark matter potential well and would be less vulnerable to stripping than near the edges of the halo. LWH also loses mass due to tidal stripping at $z\sim14$. Both haloes have a stellar mass that grows by more than a factor of 100 after they merge completely with their Superhosts at $z{\sim }8$. Prior to reaching the atomic cooling threshold and forming the black holes, both MMH and LWH were unable to form stars; the extrapolated typical stellar masses from Behroozi et al. are shown only for reference at these early redshifts (dotted curves).}
 \label{fig:stellar_mass_vs_z_beh_2019}
\end{figure}

Stellar mass is shown in  Fig.~\ref{fig:stellar_mass_vs_z_beh_2019}, where we have converted $M_{\text{peak}}$ (the peak historical halo mass) to stellar mass using the function introduced in Appendix~J of \citet{Behroozi_2019}. Though stellar mass is typically gradually increases, some stellar mass is lost during the tidal stripping event. This stellar mass is lost at a rate which is much smaller than halo mass loss rate, as the stars would concentrate near the centre of the dark matter potential well and would be less vulnerable to stripping than near the edges of the halo. This is accounted for by \citet{Behroozi_2019}, where a fixed $M_{\rm peak}$ but increasing redshift leads to a decrease in $M_*$. We note prior to reaching the atomic cooling threshold and forming the black holes, both MMH and LWH were unable to form stars; the extrapolated typical stellar masses from Behroozi et al. are shown only for reference at these early redshifts (dotted curves). This overestimate of stellar mass results in a slightly conservative $M_{\rm bh}$ {\it vs.} $M_*$ relation.

\begin{figure*}
    \includegraphics[width=\textwidth, height=0.85\textheight]{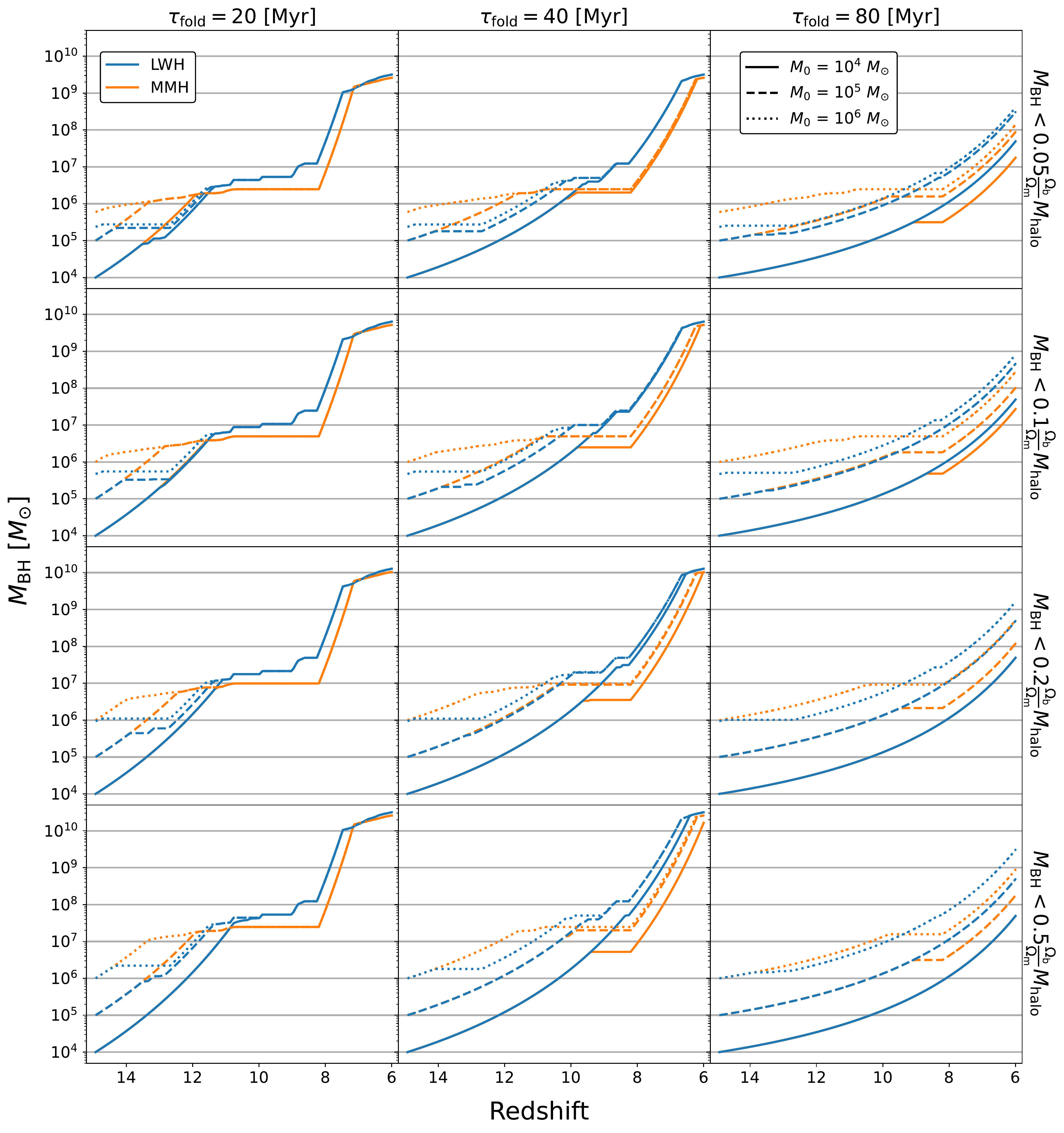}
    \caption{Black hole mass {\it vs.} redshift, assuming Eddington-limited accretion. The black hole seeds form in MMH at $z=16.4$ and LWH at $z=15.3$. We explore the initial seed masses $M_{\rm i} \in \{10^4, 10^5, 10^6\}$ ${\rm M_\odot}$. The black hole's growth is halted if its mass reaches a fraction of the halo's baryonic matter, restricting $M_{\rm bh} \leq f_{\rm bh} M_{\rm halo}\Omega_{\rm b}/\Omega_{\rm m}$ for $f_{\rm bh}  \in \{ 0.05, 0.1, 0.2, 0.5\}$, corresponding to each row. Columns represent different e-folding times $\tau_{\rm fold} \in \{20, 40,80\}$ Myr.}
    \label{fig:bh_vs_z}
\end{figure*}

Black hole growth is shown in Fig.~\ref{fig:bh_vs_z}, for the ranges of folding times $\tau_{\rm fold}$, mass caps $f_{\rm bh}$, and initial seed mass $M_{\rm i}$ mentioned in the previous section. The black hole growth starts at the first available snapshot with redshift less than the black hole's birth, $z=14.926$, meaning periods of growth before reaching the cap are slightly conservative. $\tau_{\rm fold}$ depends on the radiative efficiency factor $\epsilon$, where $\tau_{\rm fold}=40$ Myr corresponds to  $\epsilon\approx 0.1$.
We find that the final black hole mass at redshift $z=6$ ranges from $10^7 {\rm M_\odot}$ in the most strict case (top right, solid lines) to $10^{10} {\rm M_\odot}$ in the least strict case (bottom left, dashed lines). For $\tau_{\rm fold} \geq 40$ Myr, the final BH mass is roughly independent of initial mass and is instead governed by the mass cap $f_{\rm bh}$.  For $\tau_{\rm fold} = 80$ Myr, the BH does not reach the cap, and the final mass of the BH therefore scales linearly with its initial assumed mass.

\begin{figure*}
 \includegraphics[width=\textwidth, height=0.85\textheight]{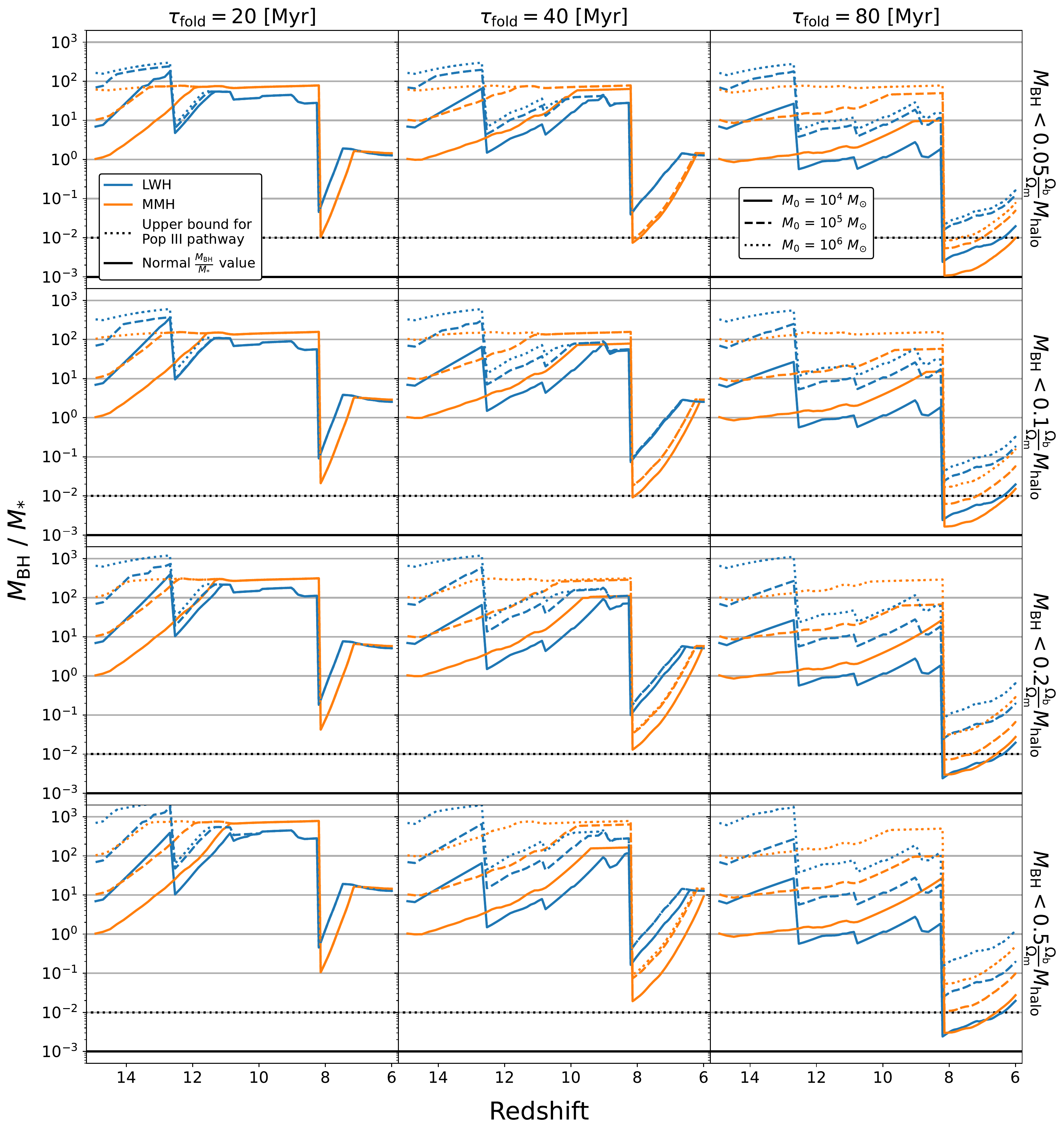}
 \caption{The ratio of black hole mass to stellar mass {\it vs.} redshift. The MMH and LWH seeds form at $z=16.4$ and $z=15.3$ respectively. The panels show the same parameter combinations as Fig.~\ref{fig:bh_vs_z}. In the most conservative panel (top right, solid lines), we find that $M_{\rm bh}/M_* \gtrsim 1$ until the mergers at redshift $z\sim8$. The solid line marks the typical black hole to stellar mass ratio for low redshift galaxies, $\sim 10^{-3}$. The dotted lines show a rough upper bound on the mass ratio for Pop III seeds, $\sim 10^{-2}$, as discussed in \S~\ref{sec:dbch_disc}.}
 \label{fig:ratio_vs_z}
\end{figure*}

The ratio of black hole to stellar mass for the range of our parameter combinations is shown in Fig.~\ref{fig:ratio_vs_z}. We also show the approximate upper bound on this ratio in the Pop III pathway ($10^{-2}$) and the approximate typical value in low redshift galaxies ($10^{-3}$). In all cases, we have $M_{\rm bh}/M_* \geq 1$ initially, at redshift $z=14.926$. For most parameters, the ratio remains $M_{\rm bh}/M_* >1$ during most of the black hole's life until both haloes merge completely with their respective Superhosts at $z\sim8$. Even in the most conservative case (top right panel with ${\rm M}_i = 10^4 {\rm M_\odot}$), this mass ratio is orders of magnitude above the value for the Pop III pathway before this merger.

\subsection{The mass relation in the DCBH pathway}
\label{sec:dbch_disc}

For nearly every parameter combination that we have explored, both haloes have a black hole to stellar mass ratio of  $M_{\rm bh}/M_* \gtrsim 1$ from the time the black hole seed is born until redshift $z\sim8$, when the haloes merge completely with their respective Superhosts. This is true even for the "worst case", when the black hole growth is assumed to be the slowest ($\tau_{\rm fold} = 80$ Myr), for the smallest initial mass ($M_{\rm i} = 10^4\ {\rm M_\odot}$), and using the tightest mass cap ($f_{\rm bh} = 0.05$). This strict scenario is shown by the solid lines in the top right panel of Fig.~\ref{fig:ratio_vs_z}.

As an alternative to the DCBH pathway, Pop III stars could create the seeds which then grow via periods of super-Eddington accretion and mergers into the $10^9\ {\rm M_\odot}$ black holes that we observe at high redshift today. \citet{Agarwal_2013} proposed that this formation pathway is distinguishable from the heavy-seed pathway via this mass ratio, since it is much higher for heavy seeds. \citet{Habouzit_2017} have shown that the Pop III pathway results in $M_{\rm bh}/M_* < 10^{-2}$, due primarily to strong supernova feedback which limits the black hole growth at early stages. \citet{Habouzit_2022} confirm that feedback at high redshift could maintain a $M_{\rm bh}$-$M_*$ relation which is similar to the local value of $\sim 10^{-3}$.  \citet{Valiante_2018} find results that agree with this (see their Fig.~2 in particular).  

For our MMH and LWH, the mass ratio remains $M_{\rm bh}/M_{\rm stars} > 1$ until their mergers near redshift $z=8$ with their respective massive Superhost haloes. If we have an OMBG that avoids this collision, it seems likely that this ratio will be greater than unity at redshifts below $z<8$, and continue to be distinguishable from Pop III seeds which have grown to a similar mass but have a much greater accompanying stellar mass. Future work will aim to calculate the expected lifetime (or indeed, probability distribution of lifetimes) of an OMBG before it collides with a more massive Superhost and loses this unique mass ratio.  We note that this diagnostic does not distinguish DCBHs from Pop III seeds which grow rapidly via hyper-Eddington accretion into a $\sim 10^5\ {\rm M_\odot}$ black hole seed in the atomic-cooling halo (ACH), also producing an OMBG \citep{Inayoshi_2016}. The uniqueness of this mass ratio affords a relatively long-lasting window for directly detecting an OMBG and collecting evidence for the heavy seed pathway. See \S~\ref{sec:detection} for a discussion of observationally detecting this mass relation.

We note that these results are limited by modeling black hole growth via Eddington accretion, which assumes a continuous supply of dense gas. While the original Renaissance simulations were run down to redshift $z=15$, the high resolution re-simulation that hosts our DCBHs was stopped at the time of the dense protostar formation for MMH at $z=16.4$ and LWH at $z=15.3$.  As a result, subsequent gas supplies near the protostars in these haloes cannot be determined. However, as summarized in \S5.5 of \citet{Inayoshi_2020}, while most massive seed BHs would hardly grow due to feedback, seeds born in overdense regions like the regions that host our two DCBHs have been found in other simulations to grow efficiently via intense cold accretion streams.

Even for a fixed $\tau_{\rm fold} = 80$ Myr, this mass relation stays well above the standard value for all other parameters explored. This suggests that a DCBH which goes through periods of accretion (with $\tau_{\rm fold} < 80$ Myr) and starvation would likely maintain an outstanding mass relation until a merger with a much more massive halo. If growth is slower than this, our DCBHs would not grow into the ${\sim}10^9 {\rm M}_\odot$, but this model shows that, in general, a DCBH which does become a SMBH at redshift $z \sim 6$ would likely remain an outlier in the mass relation until merging with a more massive halo.

\subsection{An alternative model for BH growth}
\label{sec:discussion_growth}

We have so far assumed  Eddington-limited black hole growth, though mass inflow rates could permit super-Eddington accretion. \citet{Hu_2022a} and \citet{Hu_2022b} have recently explored BH accretion when the large-scale feeding rate is substantially higher than $\dot{M}_{\rm Edd}[\equiv L_{\rm Edd}/(\epsilon c^2)]$ in radiation-hydrodynamics simulations. They find, consistent with earlier results (\citealt{Jiang_2014, Sadowski_2015, Inayoshi_2016, Toyouchi_2021}), that when the external gas supply rate $\dot{M}_{0}$ exceeds the Eddington rate, photons trapped in the dense flow produce strong outflows which then decrease the mass inflow rate with distance $r$ from the black hole as $\propto r^p$ with $p\sim 0.5$ - $0.7$. They provide a simple prescription for the rate of black hole growth $\dot{M}_{\rm bh}$ (see their Eq.~1) which simplifies to
\begin{align}
    \dot{M}_{\rm bh} = \dot{M}_{0}^{1-p} \left(\frac{3}{5}\dot{M}_{\rm Edd}\right)^p \ \ \ \ \ \ \ \text{if } \ \frac{3}{5}\dot{M}_{\rm Edd} \leq \dot{M}_{0}
\end{align}
and $\dot{M}_{\rm bh}=\dot{M}_{0}$ otherwise. We approximate gas supply as constant,
\begin{align}
\dot{M}_{0} \approx \mathscr{F}\frac{\Omega_{\rm b}}{\Omega_{\rm m}}\frac{M_{\rm halo}(t) - M_{\rm halo}(t_0)}{t-t_0}
\end{align}
for $\mathscr{F} \sim 0.1$. This allows us to solve for the mass of the black hole,
\begin{align}
    M_{\rm bh}(t) = \dot{M}_{0} \left(\frac{3}{5\tau_{\rm fold}}\right)^{\frac{p}{1-p}}[(t-t_0) (1-p)]^{\frac{1}{1-p}} + M_{\rm bh}(t_0)
\end{align}
for a period from $t_0$ to $t$ where $(3/5)\dot{M}_{\rm Edd} \leq \dot{M}_{0}$, otherwise $M_{\rm bh}(t) = \dot{M}_{0}(t-t_0) + M_{\rm bh}(t_0)$. Using $p=0.6$, the resulting mass for this model of BH growth and the ratio of mass calculated between this and our original model are shown in Fig~\ref{fig:bh_vs_z_super} and Fig~\ref{fig:bh_vs_z_super_ratio}, respectively. The updated values for $M_{\rm bh}/M_*$ are shown in Fig~\ref{fig:bh_over_M_stars_vs_z_super}.

\subsection{An alternative stellar mass calculation}
\label{sec:discussion_2}

\begin{figure}
 \includegraphics[width=\columnwidth]{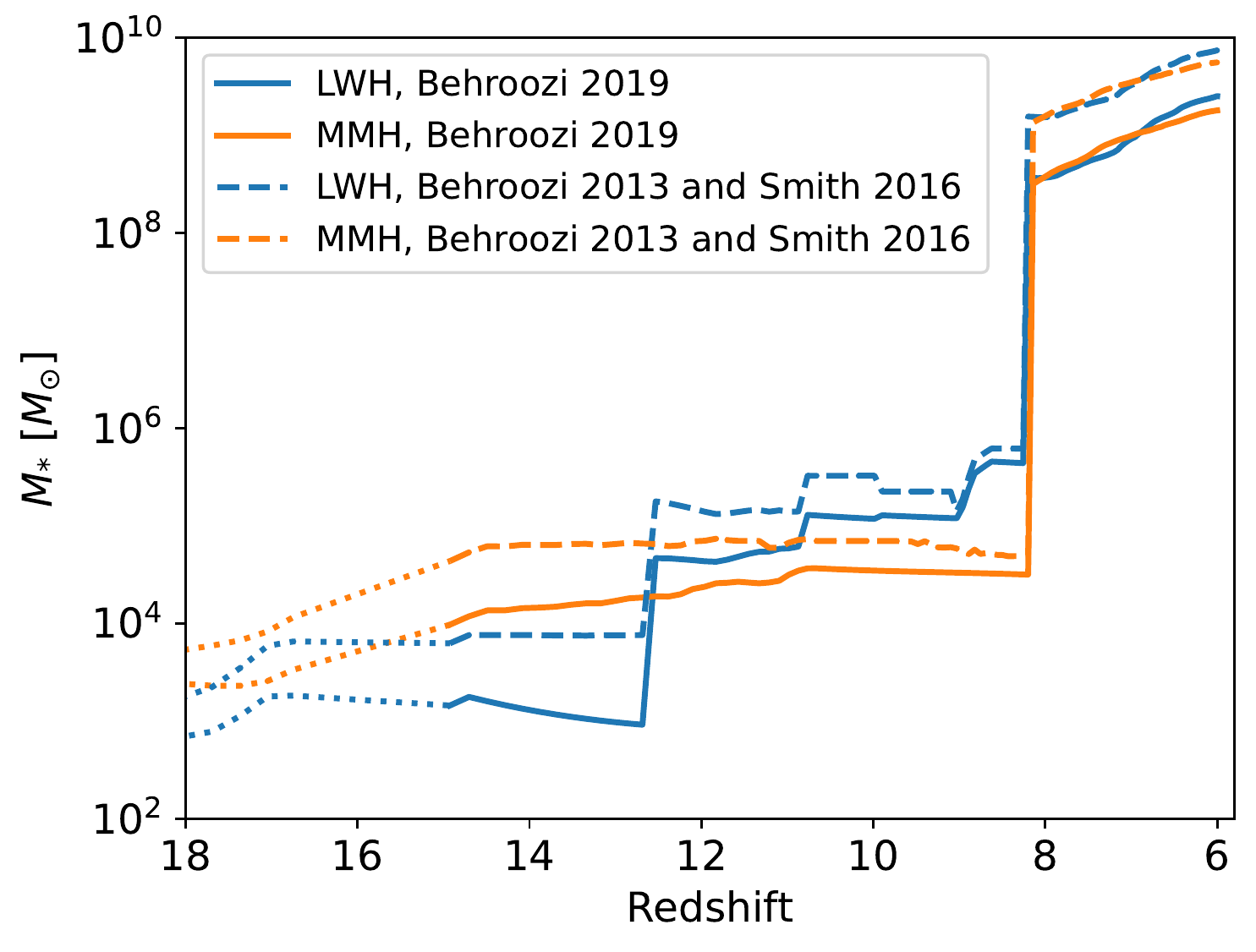}
  \includegraphics[width=\columnwidth]{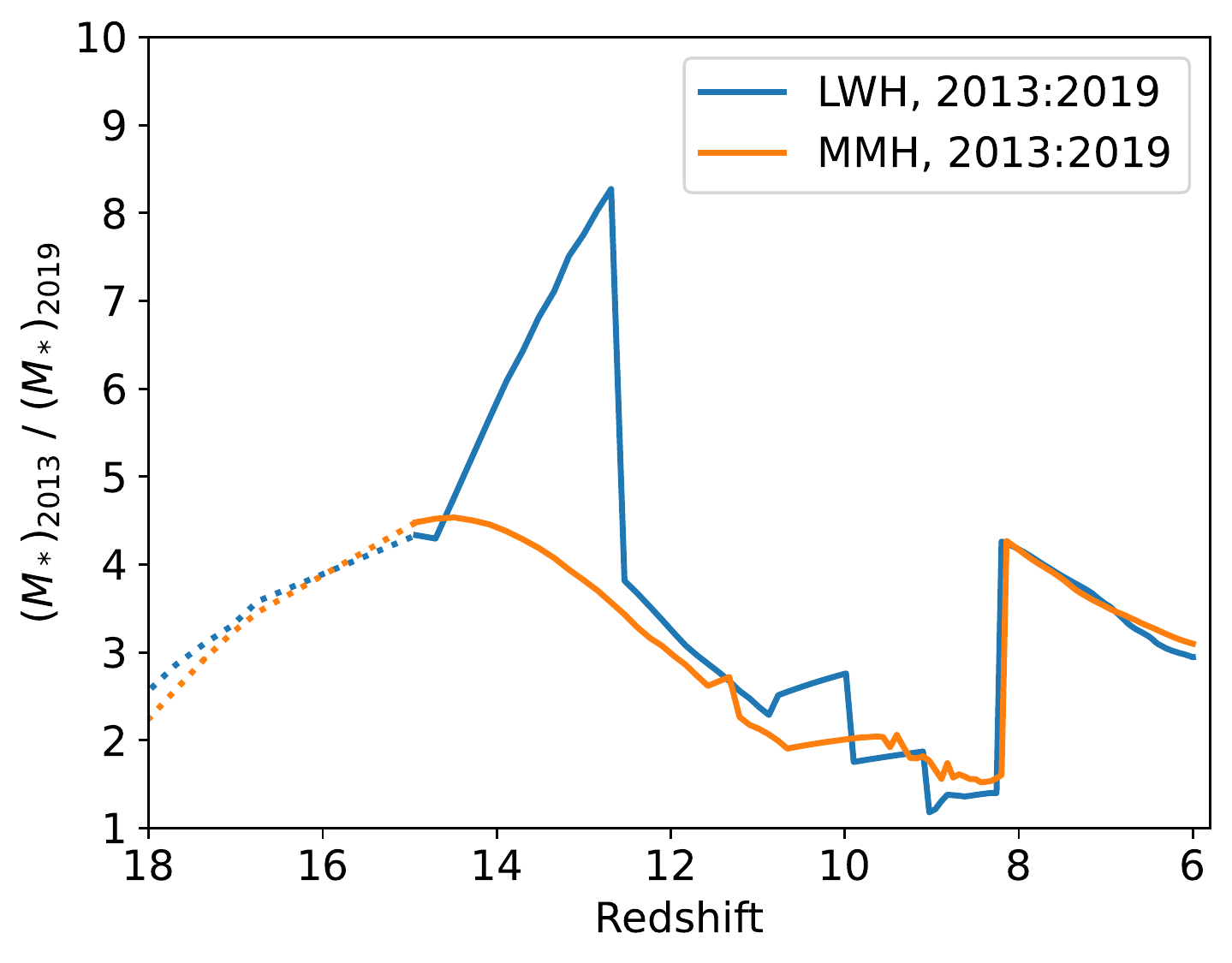}
\caption{{\it Top:} Comparing two different stellar mass calculations. The solid lines are calculated using Eq.~J1 of \citet{Behroozi_2019}, which is a function of the peak historical halo mass, $M_{\rm peak}$, and constants adopted from row 15 of their Table~J1. See \S \ref{sec:methodsSM} for a discussion on the choice of these constants. The dashed lines are calculated using Eq.~3 of \citet{Behroozi_2013}, which is a function of the instantaneous halo mass, and periods of tidal stripping (where $\dot{\rm M}_*<0$) are calculated separately using Eq.~1 from \citet{Smith_2016}. The stellar masses should be considered unreliable at the highest redshifts, where both stellar-mass determinations are well outside of their calibrated range. Prior to reaching the atomic cooling threshold and forming the black holes, both MMH and LWH were unable to form stars; the extrapolated typical stellar masses from Behroozi et al. are shown only for reference at these early redshifts (dotted curves). {\it Bottom:} The ratio of these two different stellar masses. The spike in the stellar mass ratio for LWH near redshift $z=12$ is due to significantly more mass being preserved during tidal stripping when calculated with the \citet{Smith_2016} prescription.}

 \label{fig:stellar_mass_vs_z_comparison}
\end{figure}

The BH to stellar mass ratio is heavily dependent on our BH and stellar mass "painting" method. When considering other methods, we find that the ratio is well above the typical value of $10^{-3}$ in all cases (pre-merger). Our stellar mass calculation following \citet{Behroozi_2019} uses the largest historical halo mass, $M_{\rm peak}$, where a fixed peak but increasing redshift will result in a decreasing mass for our halo mass ranges. This allows us to lose some stellar mass due to tidal stripping, although indirectly. An alternative to this is to calculate the stellar mass using the instantaneous halo mass at all redshifts, then use a tidal stripping formula during periods of halo mass loss. During periods of mass loss, halo mass loss fractions are much higher than stellar mass loss fractions, due to stellar mass concentrating at the centre of the halo and being more resistant to the stripping. This is why calculating stellar mass as a function of instantaneous halo mass is unreliable for periods of tidal stripping. For the purposes of illustration, we nevertheless compare our stellar mass to this alternative, where Eq.~3 of \citet{Behroozi_2013} gives stellar mass as a function of instantaneous halo mass and Eq.~1 from \citet{Smith_2016} gives the fraction of stellar mass loss as a function of the fraction of DM halo mass loss, $f_{\rm str} = 1-e^{-\alpha f_{\rm DM}}$. Our results use their best-fit value of $\alpha = 14.2$. We explored several different values of $\alpha$ corresponding to how extended the galaxy stellar component is in comparison to the dark matter halo (their Eqs.~2-5), and found negligible differences.  Our comparison is shown in Fig.~\ref{fig:stellar_mass_vs_z_comparison}, with the top panel showing the stellar mass in each calculation and the bottom panel showing the ratio of the two. Both stellar mass calculations are calibrated for halo masses of $10^{10} {\rm M_\odot}$ to $10^{15} {\rm M_\odot}$ where extrapolation can be considered reliable for regions slightly outside of this range. 

The method based on combining \citet{Behroozi_2013} and \citet{Smith_2016} yield stellar masses which are consistently above those based on \citet{Behroozi_2019}. We see significant divergences at high redshift $z>18$, though the disagreement is expected since both functions are well outside of their calibrated ranges. We see some disagreement for redshift $z<18$, though these are never an order of magnitude greater in the case of the \citet{Behroozi_2013} + \citet{Smith_2016} approach. Even for the most conservative black hole growth model, this would still lead to $M_{\rm bh}/M_*>10^{-1}$ at all redshifts before the merger (solid lines in the top right panel of Fig.~\ref{fig:ratio_vs_z}). This means our conclusion remains the same: the mass ratio of the heavy-seed pathway is distinguishable from the light-seed pathway, with the most conservative mass ratio still an order of magnitude greater than the upper bound for light seeds.

\subsection{DCBH seed density and detection}
\label{sec:detection}

W19 identified 11 metal-free ACHs in the 133.6 (comoving) ${\rm Mpc}^3$ {\it Rarepeak} region that had not hosted any prior star formation. Not all of these are heavily irradiated or suffer unusually rapid dynamical heating, meaning it is not necessary that they will all experience large mass inflow rates and become OMBGs. Furthermore, {\it Rarepeak} is $\sim1.68$ times denser than the cosmic mean and is not representative of the universe. While this may seem to put a tight upper bound on the DCBH number density, \citet{Chon_2020}, \citet{Tagawa_2020} and \citet{Regan_2020} have recently shown that the metal-free condition is not strictly necessary for the formation of SMSs -   rapid inflow rates can arise in the presence of some modest metal pollution. Accounting for metal-enriched regions that achieve the required inflow rates of $\sim0.1 {\rm M_\odot}~{\rm yr}^{-1}$ via other ${\rm H_2}$ suppression mechanisms and allow SMS formation, \citet{Regan_2020} calculates a DCBH seed number density of 0.26 (comoving) ${\rm Mpc}^3$ in the \texttt{Renaissance} simulation. After accounting for the rarity of the simulated over-density, they obtain a global average DCBH seed space density of $\sim10^{-5}$ (comoving) ${\rm Mpc}^{-3}$. This is many orders of magnitude above the observed number density of SMBHs at $z=6$, $\sim 1$ (comoving) ${\rm Gpc}^{-3}$. This means that the direct-collapse pathway could possibly account for most or all of the SMBHs, motivating future work to focus on detecting this unique mass relation.

Combining X-ray and infrared observations could establish the SMBH's location in the $M_{\rm bh}/M_*$ relation. X-ray observations could be used to detect the central black hole. \citet{Pacucci_2015} uses \texttt{CLOUDY} \citep{Ferland_2013}, a spectral synthesis code, to generate time-dependent spectra of an accreting $10^5\ {\rm M_\odot}$ black hole (in a halo of $10^8\ {\rm M_\odot}$) as a function of the matter distribution from radiation-hydrodynamic simulations and the irradiation spectrum at the inner boundary. They find a spectrum which is dominated by the infrared-submm ($1-1000 $ $\mu$m) and X-ray ($0.1-100$ keV) bands. They show that in their standard (non-slim disc) Eddington-limited accretion model the luminosity of the DCBH grows until peaking at 115 Myr, allowing {\it Athena} to make an X-ray detection ($3\sigma$ with $3{\times}10^5$s integration time) after the DCBH's first $\sim 100$ Myr, just before reaching peak luminosity, at $z=9$. This model assumes accretion within 10 pc of the black hole until gas depletion. The time dependence of the luminosity results in {\it Athena} having a detection window for $\sim 25\%$ of the total accretion time. 
{\it Lynx} is a concept being studied for a next-generation X-ray observatory~\citep{lynx} to improve on both the angular resolution and the sensitivity of {\it Athena}. At its current design, it would detect $10^4\ {\rm M_\odot}$ BHs near redshift $z=10$ and $10^5\ {\rm M_\odot}$ BHs at redshift $\gtrsim 15$. Such sensitivity could constrain the evolution of SMBHs, which would help distinguish the light {\it vs.} heavy seed pathways~\citep{Haiman_2019}.

{\it JWST} is likely able to detect star forming galaxies out to $z\sim 20$, but certainly to $z\sim 15$. {\it JWST}'s NIRCam is capable of detecting $m=30.5$ at $5\sigma$ with an ultra-deep exposure of $\sim88$ hrs \citep{Finkelstein_2015}. Similarly, \citet{Zackrisson_2011} predict that {\it JWST} should be able to detect Pop III galaxies with $M_* \sim 10^5\ {\rm M_\odot}$ and metal-enriched galaxies with $M_* \sim 10^6\ {\rm M_\odot}$ at $z\approx 10$ in ultra deep exposures, ($10\sigma, 100$ hr). Depending on the models used, this enables the detection of stellar mass $M_* \sim 10^{5-6}\ {\rm M_\odot}$ at $z \gtrsim  10$ \citep{Pacucci_2019}.

While it is unlikely that an ultra-deep {\it JWST} field will be chosen to target an X-ray candidate source, there is a chance that a detectable X-ray source will be found in these fields. Alternatively, shorter exposures could allow follow-ups to these X-ray sources. The CEERS \citep{CEERS} program will detect sources down to $m\sim 29$ with $2{,}800$s exposures and the JADES \citep{Bunker_2019} program will detect down to $m=29.8$ in $2.5 \times 10^4$ s deep field exposures. These shorter exposure times will raise the stellar mass detection threshold. Since the flux detection threshold with NIRCam for faint sources scales\footnote{See the JWST Exposure Time Calculator at https://jwst-docs.stsci.edu/jwst-near-infrared-camera/nircam-predicted-performance/nircam-imaging-sensitivity} approximately as $\propto\sqrt{t}$ with integration time $t$, {\it JWST} should be able to detect haloes with a stellar mass of $\sim 10^6 - 10^7\ {\rm M_\odot}$. Even if we assume a stricter threshold of $\sim 10^7 - 10^8\ {\rm M_\odot}$, we find that the BHs in our MMH and LWH haloes are $\geq 10^6\ {\rm M_\odot}$ for most of our growth parameters near redshift $z=10$, which results in a  $M_{\rm bh}/M_*$ ratio of $10^{-1} - 10^{-2}$. This is well above the typical low-redshift values and still above those expected in the Pop III pathway. Further, a confirmed X-ray source BH detection, together with the absence of a stellar mass detection from {\it JWST} will place a strong lower limit, $\geq 10^{-1}$, on this mass ratio, providing evidence for OMBGs. See \citet{Pacucci_2019} or  \citet{Inayoshi_2020} for more complete reviews on methods of DCBH detection.

\begin{figure}
 \includegraphics[width=\columnwidth]{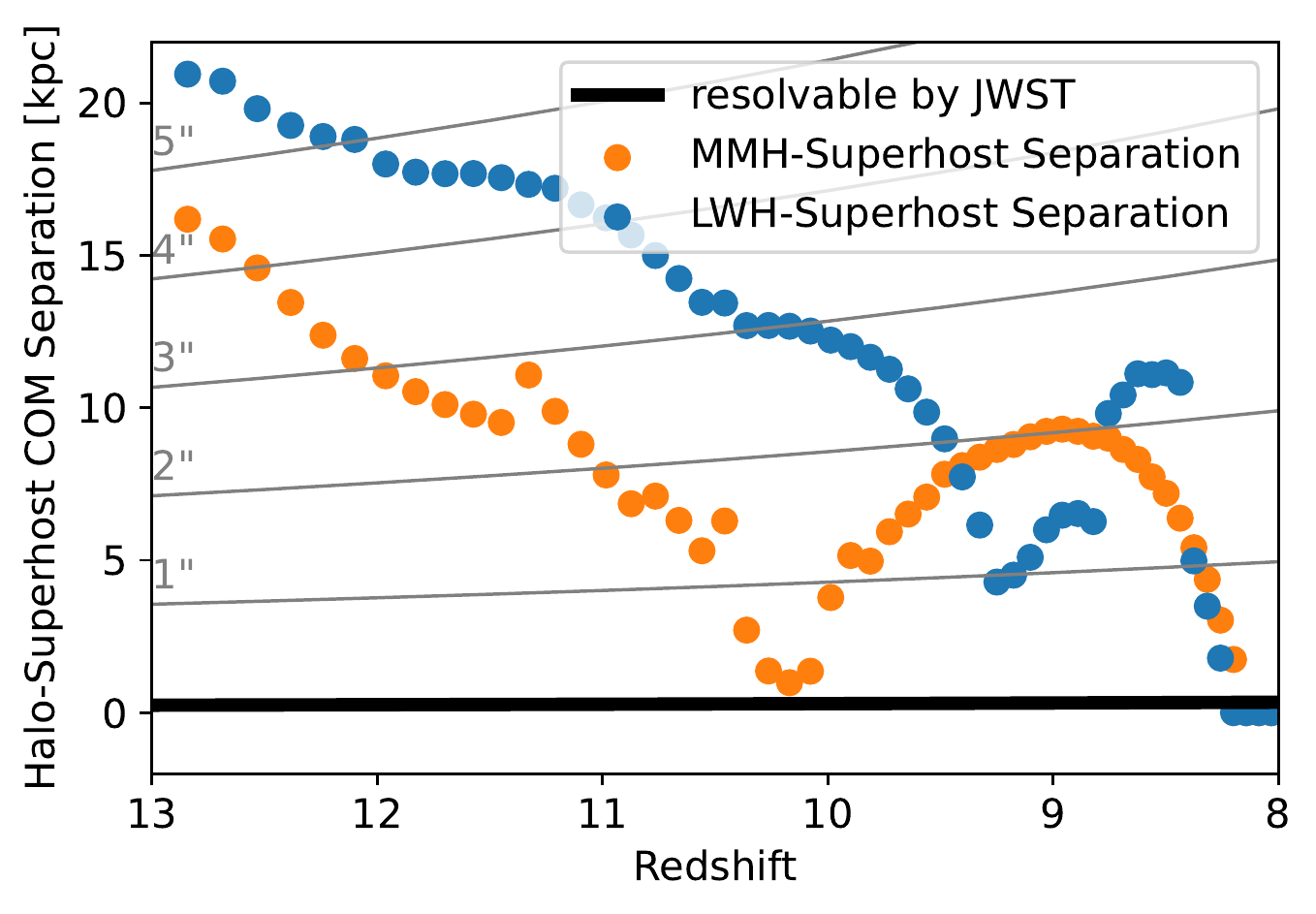}
 \caption{The separation between the centre of mass of the MMH/LWH and their Superhosts.  {\it JWST} has an angular resolution of $0.1"$, corresponding to $\gtrsim 0.5$ kpc at redshift $z\approx 10$. This minimum resolution is shown by the thick nearly horizontal black line. We find that the MMH/LWH are sufficiently far away from the centre of their Superhost halo, including even after the merger (when they become subhaloes, near redshift $z=10.7$), allowing {\it JWST} to spatially resolve them down to the redshift $z\approx8$ when the mergers are completed.}
 \label{fig:halo_parent_sep}
\end{figure}

\begin{figure*}
 \includegraphics[width=\textwidth, height=0.9\textheight]{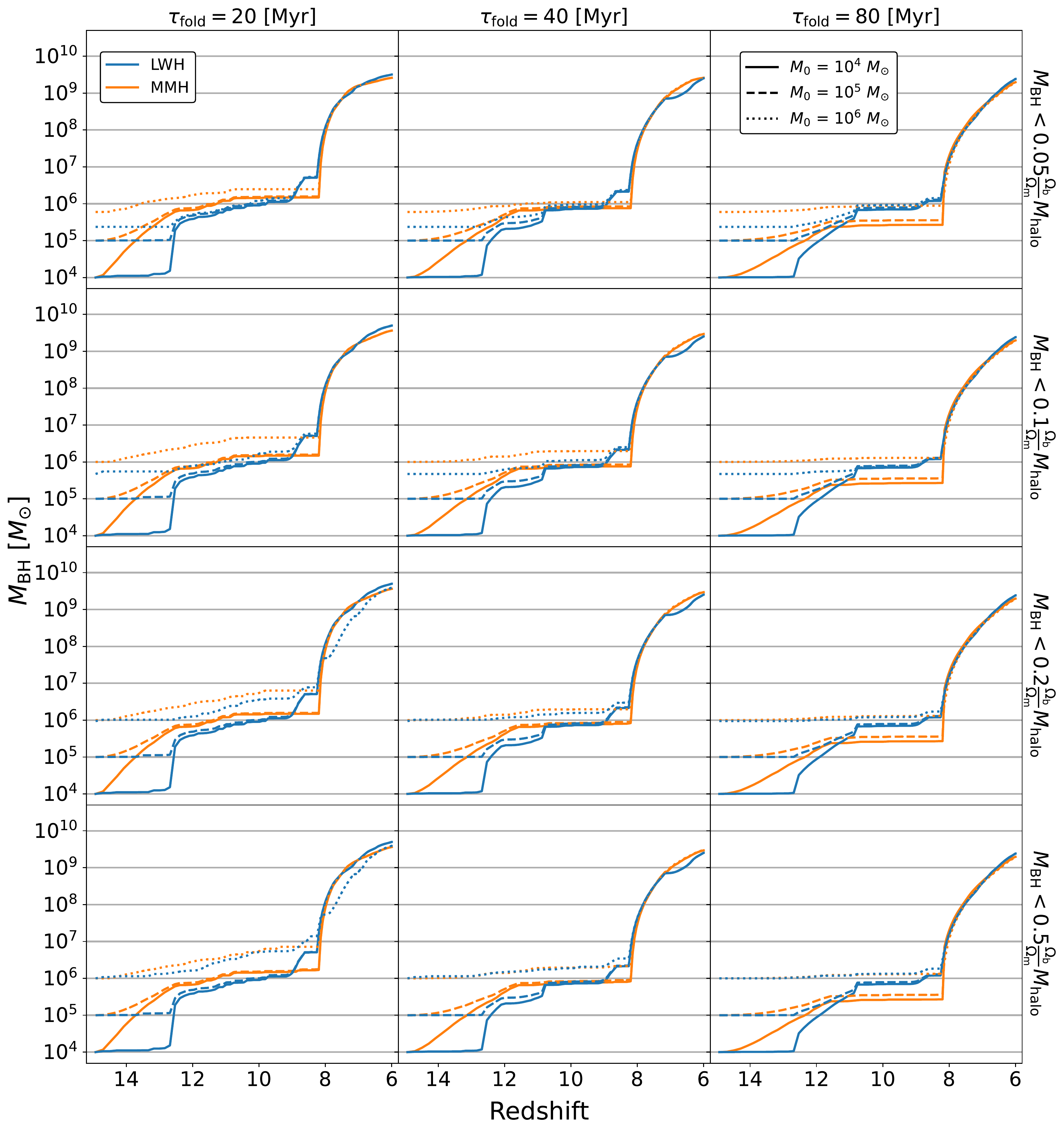}
 \caption{Black hole growth allowing super-Eddington accretion as discussed in \S~\ref{sec:discussion_growth}. The panels show the same parameter combinations as in Fig.~\ref{fig:bh_vs_z}. The black hole growth is determined by the large-scale mass inflow rate $\dot{M}_0$, where $\dot{M}_{\rm bh} = \dot{M}_{0}^{0.4} \left[(3/5)\dot{M}_{\rm Edd}\right]^{0.6}$ if $\dot {M}_{0} \geq (3/5)\dot{M}_{\rm Edd}$ (accounting for suppression due to outflows produced by trapped radiation), and $\dot{M}_{\rm bh} = \dot{M}_{0}$ otherwise.}
 \label{fig:bh_vs_z_super}
\end{figure*}

\begin{figure*}
\includegraphics[width=\textwidth, height=0.9\textheight]{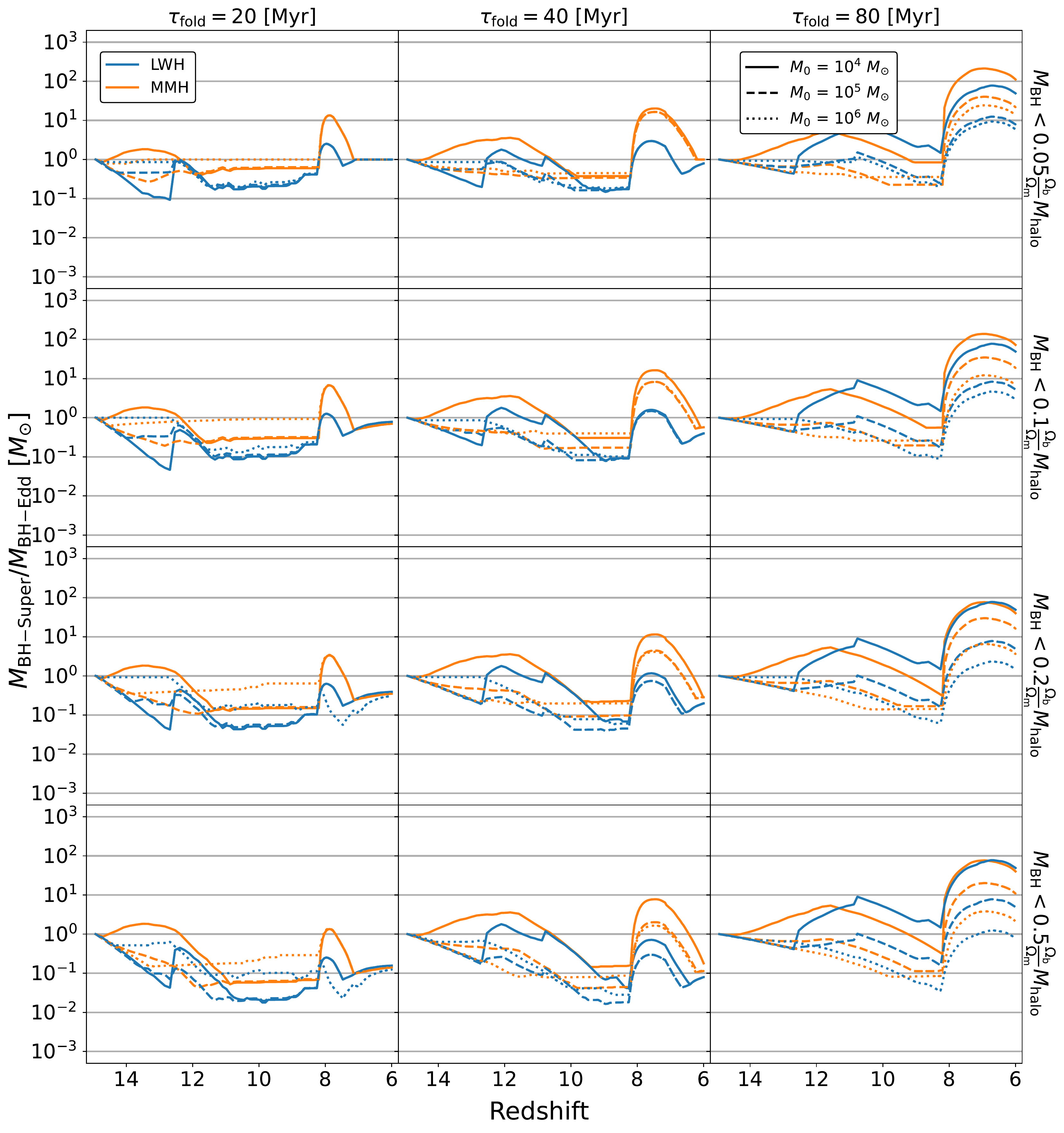}
\caption{The ratio of black hole masses in the super-Eddington growth models {\it vs.} our original Eddington-limited growth models. The panels show the same parameter combinations as in Fig.~\ref{fig:bh_vs_z}.}
 \label{fig:bh_vs_z_super_ratio}
\end{figure*}

\begin{figure*}
 \includegraphics[width=\textwidth, height=0.9\textheight]{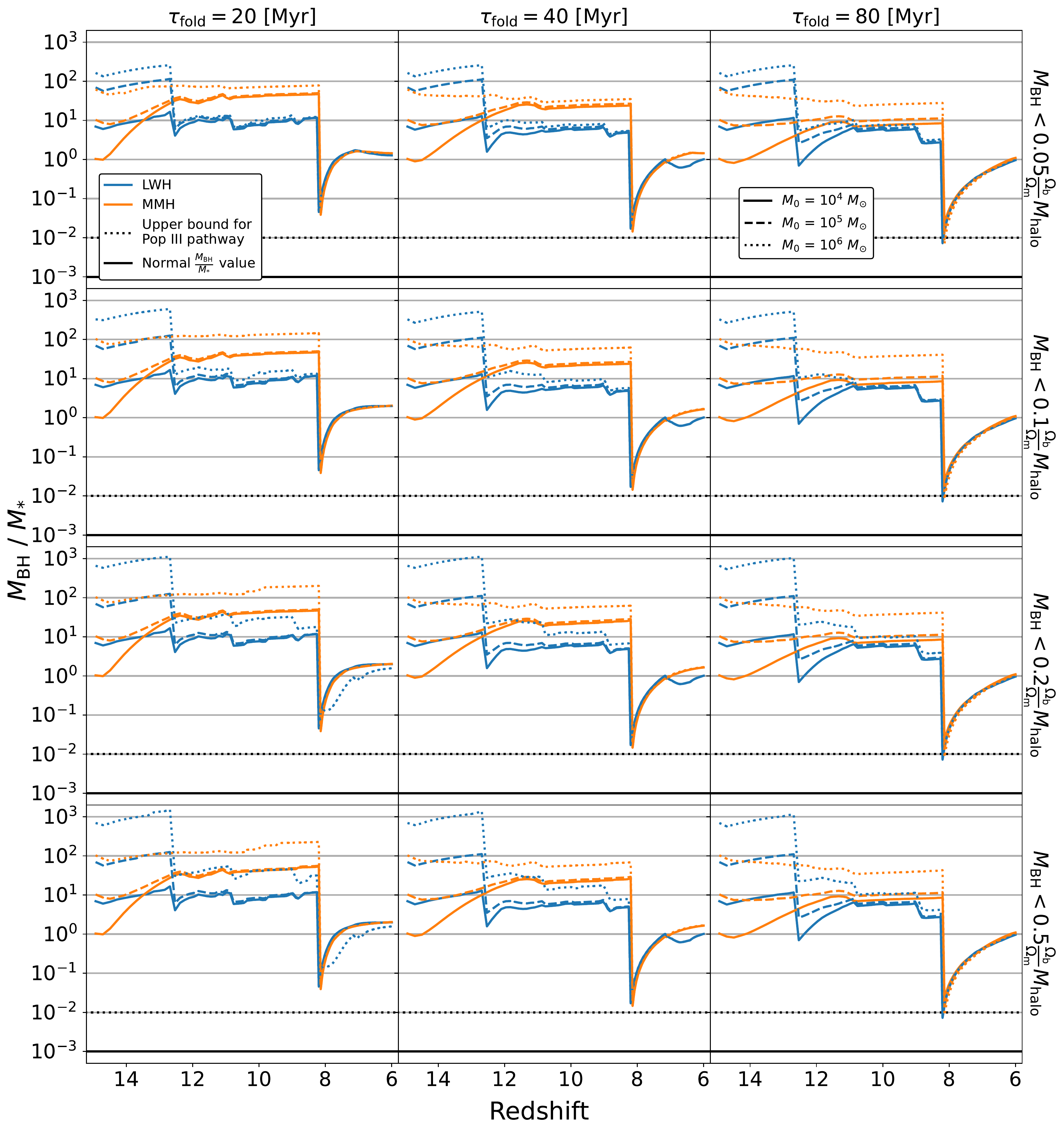}
 \caption{Updated $M_{\rm bh} {-}M_*$ ratios allowing for super-Eddington black hole growth. The panels show the same parameter combinations as in Fig.~\ref{fig:bh_vs_z}. For every combination of initial mass and folding time, the mass ratio is initially much larger than the upper bound for the light seed pathway of $10^{-2}$. The ratio approaches $10^{-2}$ during a merger with a much larger Superhost at $z\sim8$, then returns to being an outlier after rapid BH growth.}
 \label{fig:bh_over_M_stars_vs_z_super}
\end{figure*}

In {\it Renaissance}, both haloes closely interact with a more massive neighboring halo, a 'Superhost', which eventually becomes the host of the DCBH after MMH and LWH merge with them. Prior to merging completely, the distance between the DCBH-hosting haloes and their Superhosts could pose an angular resolution issue if they are too close. Fig.~\ref{fig:halo_parent_sep} shows the COM separation between our DCBH-hosting target haloes and their Superhosts, where we find they are separated by $\sim1{-}20$ kpc at all times before the merger is complete. This separation is much greater than the angular resolution of $0.1"$ of {\it JWST}, which corresponds to $\sim0.5$ kpc at redshift $z \leq 8$. MMH and LWH merge completely with their distinct Superhosts at redshift $z =8.198$ and $z=8.256$, respectively, each growing over two orders of magnitude in both halo mass and stellar mass. After their mergers, our targets become spatially unresolvable from their massive Superhosts, meaning that this information on their origin is lost.

\subsection{Mass relation {\it vs.} other detection methods}
Another distinguishing signature of the DCBH pathway is discussed in \citet{Johnson_2011}, where the ratio of the luminosity emitted in the He ${\rm II} \ \lambda 1640$ {\it vs.} the H$\alpha$ line is $\sim 2$ for the first $\sim 2$ Myr, up to an order of magnitude larger than the Pop III pathway which has $L_{1640}/L_{\rm H\alpha} \sim 0.1{-}1$ \citep[][see also \citealt{Tumlinson_2000} and \citealt{Oh_2001} for earlier proposals to use this ratio as a probe of Pop~III stars and accreting BHs at high redshift]{Johnson_2009}.  Though this luminosity ratio is a potential diagnostic tool, the mass ratio diagnostic studied here has the benefit of being several orders of magnitude larger in the DCBH pathway than in the Pop III pathway, removing potential ambiguity. Further, our target haloes have unique mass relations which remain outliers for several million years, while it is not clear how long the emission line luminosity ratio remains unique after the first few million years.

DCBHs also allow detection via their unique spectral shapes \citep{Nakajima_2022}. Both the spectral lines and continuum have features which would be unique to the DCBH scenario. The spectra found in \citet{Pacucci_2015} have a steep slope in the infrared, due to radiation from the DCBH being reprocessed at lower energies by intervening matter \citep{Pacucci_2016}. \citet{Inayoshi_2022} have recently found that Balmer lines for black holes accreting at super-Eddington rates are $\sim 2-7$ times stronger than in low-$z$ quasars. This is because the gas is denser and has a larger column density around these DCBHs than in a usual thin disk, and so excitation from $n$=2 (populated by trapped Ly$\alpha$) to $n\geq3$ states (via collisions) is more common.  Additionally, unusually strong broad OI lines are predicted, as a result of Ly$\beta$ fluorescence. For further review, see \citet{Ricarte_2018}, where the authors discuss probes that distinguish heavy seed from light seed pathways, including differences in SMBH occupation fractions and gravitational wave event signatures.

\section{Conclusions}
\label{sec:conclusion}

The direct-collapse pathway remains a promising explanation for the origin of SMBHs of mass $>10^9\ {\rm M_\odot}$ at redshift $z \geq 6$. Future work should therefore aim to distinguish between the DCBH pathway and alternatives. The idea to use a uniquely large $M_{\rm bh}/M_*$ ratio to differentiate between the Pop III pathway was proposed in \citet{Agarwal_2013}. \citet{Habouzit_2017, Habouzit_2022} and \citet{Valiante_2017, Valiante_2018} confirmed that this ratio is much higher in the heavy seed pathway than the light seed pathway. \citet{Visbal_2018} then investigated how long DCBHs in these overly massive black hole galaxies (OMBGs) remain outliers in this relation, but only for the first $\sim 100$ Myr of the seed's existence, and for random atomic cooling haloes.

Utilizing the \texttt{Renaissance} simulation data and focusing on two target haloes identified by W19, we have shown that this ratio remains $M_{\rm bh}/M_* > 1$ for $\sim 500$ Myr;  a value much larger than expected in light-seed pathways. While this work shows that growth via minor mergers maintains this mass relation for these two OMBGs, a forthcoming paper will use Monte Carlo merger trees to calculate the expected lifetimes of ${\sim 10^4}$ OMBGs before they merge completely with more massive ${\gtrsim} 10^{11}\ {\rm M_\odot}$ Superhosts, which causes $M_{\rm bh}/M_*$ to approach the standard low-redshift value of $\approx 10^{-3}$. This approach will put better constraints on the expected number of OMBGs as a function of redshift, setting an upper bound for their number density. \citet{Valiante_2017} explored a similar idea (see their Fig.~2) and found that it is very rare for a heavy-seed hosting halo to last more than $\sim 100$ Myr before experiencing a major or minor merger, though we have shown that in the case of MMH and LWH, minor mergers maintain this unusual mass relation.

MMH and LWH merge with their respective Superhosts near redshift $z\sim11$. We find that both haloes remain spatially resolvable from these more massive Superhosts until redshift $z\sim8$, when their mergers are complete and they are no longer distinguishable. With a combination of infrared observations from {\it JWST} and X-ray observations from {\it Athena} and/or {\it Lynx}, there is promise that we can directly detect this unique mass relation, which would provide strong evidence in favor of the DCBH formation pathway.

Our exploration has assumed a DCBH pathway. However, there are other viable mechanisms where a heavy seed can form in an atomic-cooling halo with comparably low stellar mass, such as a $ 10^5\ {\rm M_\odot}$ seed via hyper-Eddington accretion or through the collapse of a dense stellar cluster (see Introduction). We note that this would likely produce an OMBG, meaning it would be indistinguishable from a DCBH on the basis of the mass ratio diagnostic proposed here. This diagnostic therefore distinguishes between heavy seeds and light seeds with feedback-limited accretion.  

\section*{Acknowledgements}
We thank Eli Visbal and Kohei Inayoshi for useful comments on an earlier version of our manuscript. ZH thanks Raymond Dudley for some initial work on the topic of this paper. ZH acknowledges support from NASA grant NNX17AL82G and NSF grants AST-1715661 and AST-2006176. JHW acknowledges support from NASA grants NNX17AG23G, 80NSSC20K0520, and 80NSSC21K1053 and NSF grants OAC-1835213 and AST-2108020. The \texttt{Renaissance} simulation was performed on Blue Waters operated by the National Center for Supercomputing Applications (NCSA) with PRAC allocation support by the NSF (awards ACI-0832662, ACI-1238993, ACI-1514580). Subsequent resimulations and analysis were performed with NSF’s XSEDE allocation AST-120046 and AST-140041 on the Stampede2 resource. The freely available code \texttt{yt} \citep{yt} and plotting library matplotlib \citep{Hunter_2007} were used to construct the plots in this paper. Tree analysis was performed with \texttt{ytree} \citep{ytree}.  Computations described in this work were performed using the publicly-available code \texttt{Enzo}, which is the product of a collaborative effort of many independent scientists from numerous institutions around the world.

\section*{Data Availability}
The \texttt{Renaissance} simulation data are available in the Renaissance Simulations Laboratory, at https://rensimlab.github.io/. The code used during the preparation of this manuscript is available at this \href{https://github.com/mscoggs/dcbh_in_renaissance}{github repository}.  All other data will be shared on reasonable request to the corresponding author.

\bibliographystyle{mnras}
\bibliography{references} 
\bsp
\label{lastpage}
\end{document}